\documentclass[twocolumn]{revtex4}

\def\be{\begin{equation}}
\def\ee{\end{equation}}
\def\la{\label}
\def\bea{\begin{eqnarray}}
\def\eea{\end{eqnarray}}
\def\non{\nonumber}
\def\ci{\cite}
\def\la{\label}
\def\bib{\bibitem}

\def\le{\left}
\def\ri{\right}

\def\s8{\sigma_8}

\def\fr{\frac}
\def\pp{\partial}
\def\pu{\pp_\mu}
\def\pU{\pp^\mu}

\def\non{\nonumber}

\usepackage{graphicx}

\begin{document}

\title{The Mass, Normalization and Late Time behavior of the Tachyon Field}

\author{A. de la Macorra, U. Filobello}
\affiliation{Instituto de F\'{\i}sica, UNAM, Apdo. Postal 20-364,
01000 M\'exico D.F., M\'exico}
\author{G. Germ\'an}
\affiliation{Centro de Ciencias F\'{\i}sicas, Universidad Nacional
Aut{\'o}noma de M{\'e}xico, Apartado Postal 48-3, 62251 Cuernavaca,
Morelos,  M{\'e}xico}

\begin{abstract}

We study the dynamics of the tachyon field $T$. We derive the mass
of the tachyon as the pole of the propagator which does not coincide
with the standard mass given in the literature in terms of the
second derivative of $V(T)$ or $Log[V(T)]$. We determine the
transformation of the tachyon in order to have a canonical scalar
field $\phi$. This transformation reduces to the one obtained for
small $\dot T$ but it is also valid for large values of $\dot T$.
This is specially interesting for the study of dark energy where
$\dot T\simeq 1$. We also show that the normalized tachyon field
$\phi$ is constrained to the interval $T_2\leq T \leq T_1$ where
$T_1,T_2$ are zeros of the original potential $V(T)$. This results
shows that the field $\phi$ does not know of the unboundedness of
$V(T)$, as suggested for bosonic open string tachyons. Finally
we study the late time behavior of tachyon field using
the L'H\^{o}pital rule.

\end{abstract}


\maketitle

\section{Introduction}

In recent times a great amount of work has been invested in studying
the dynamics of tachyon field \ci{DB2},\ci{slowroll}. This field is motivated from
string-brane physics and represents the lowest energy level of an
unstable Dp-brane or that of a brane-antibrane system \ci{tachsen}.
One of the generic properties of this class of systems is precisely,
the existence of the tachyon field $T$.  Therefore, the
phenomenology of the tachyon is very important in understanding the
low energy limit of string-brane models.

We would like to determine some of the physical properties of the
tachyon which are important to set up the transition from
string-brane  theory to an effective low energy 4-D field theory
\ci{tachsen}. Since the tachyon field is the last surviving mode of
unstable branes, the tachyon reheating or decay processes are
specially important in the low energy theory. The usual  reheating
or decay processes  are given in terms of the coupling of the
original  field to other lighter fields (scalars or fermions) and to
its mass. Therefore, the phenomenology of the tachyon depends
heavily on its mass.

In the case of Dp-brane systems in superstrings,  the potential
$V(T)$ has been conjectured \ci{tachsen} to be tachyonic at the
origin $T=0$, have a $Z_2$ symmetry $T\rightarrow -T$, and have a
stable minimum. On the other hand for a bosonic Dp-brane systems,
the potential $V(T)$ has   been conjectured \ci{tachsen}  to be
tachyonic at the origin, have a minimum for $T=T_1>0$, with
$V(T_1)=0$, and be unbounded from below for $T<0$ with $V(T_2)=0$.
The study of bosonic tachyons has not been as exhaustive as for the
superstring case because it has been argued that the unboundedness
of the potential is a problem and because potentials with a minimum
at a finite value of $T$ are unstable \ci{tachunst}. Here we will
show that the normalized tachyon $\phi(T)$, defined below,  does not
see the unboundedness of $V(T)$ since the field $\phi(T)$ is
constrained in the interval $T_2\leq T \leq T_1$

In this work we are interested in determining the transformation
which renders the tachyon field $T$ into a standard normalized
scalar field $\phi$ and derive the  effective potential
$U(\phi(T))$. The transformation we obtain reduces to the one
obtained for small kinetic term $|\dot T|\ll 1$ but is also valid
for large values of $\dot T$. This is specially relevant for the
study of dark energy where $\dot T\simeq 1$ with $V(T)\simeq 0$.

We study the late time behavior of $T$ and $\phi$ and with the use
of the L'H\^{o}pital rule we derive some exact slow roll conditions, valid
for late time. It is then easy to obtain the scaling solution
potentials and determine if the potential leads to an accelerating universe.

This work is organized as follows. In section \ref{secT} we present
an overview of the tachyon field $T$. In section \ref{secM} we  use
the definition of the mass of a scalar field as the pole of the
propagator to determine the mass of the tachyon. In section
\ref{secN} we derive the transformation of the field $T$ in order to
have a canonical normalized scalar field $\phi$ while in section
\ref{secG} we study the model independent behavior of $\phi$ and we
 present some examples. Finally in section \ref{secL} we use the
L'H\^{o}pital rule to study the late time behavior of the scalar field and
  we conclude in section \ref{secC}.

\section{Tachyon Field $T$}\la{secT}

The string tachyon field $T$ is given in terms of  a Born-Infeld "BI" Lagrangian
\ci{tachsen},
 \be\la{L}
L=-V(T)\sqrt{1-\pu T\pU T} ,
\ee
with  $V(T)$ the potential. The potential is in principle arbitrary but
if we want to describe a stringy tachyon then
 $V$ has a maximum at the origin $T=0$  with positive energy. The evolution
of $T$ can be obtained by solving  the equation of motion of eq.(\ref{L})
 and
for an homogenous field in a Minkowski metric is given by
\be\la{eqM}
\fr{\ddot T}{1-\dot T^2} + \fr{V_T}{V}=0 ,
\ee
where $\dot T\equiv dT/dt$ and $V_T\equiv dV/dT$. From eq.(\ref{eqM}) the second term could define
an effective  potential
\be\la{F}
F(T)\equiv Log[V(T)] ,
\ee
 with $dF/dT=V_T/V$.
 The  potential $F$
has been widely used in the literature as the effective potential
but as we will  show   neither  $F$ nor $V$ are the correct
effective potential for $T$ \ci{slowroll}.

The energy density and pressure are given by
\be\la{rho}
\rho=\fr{V(T)}{\sqrt{1-\dot T^2}}, \hspace{1cm} p=L=-V(T) \sqrt{1-\dot T^2} .
\ee
In terms of $\rho$ we can write the pressure as
\be\la{p}
p=-\fr{V^2}{\rho}= w \rho ,
\ee
with the effective equation of state given by $w=-V^2/\rho^2$. From eq.(\ref{p})
we see that the tachyon field can be treated as a Chaplygin gas
for a flat potential $V$ or as quintessence with
$w=-V^2/\rho^2$.

The equation of motion (\ref{eqM}) implies that $\dot \rho =0$, i.e. the energy density is conserved,
and $\dot T$ becomes a function of $T$, which from eq.(\ref{rho}) is
\be\la{dT}
\dot T^2=1-\fr{V(T)^2}{\rho^2}.
\ee
If at the origin $T=T_i=0$ we set up the  initial condition
 $\dot T_i=0$,  then from eq.(\ref{rho}) we see that the energy density
takes the constant value $\rho=V_i\equiv V(T_i)$ and $V_i$ represents
the maximum value of the potential $V$, i.e. $V(T)\leq V_i$ for all $T$.
Furthermore,
if the minimum of the potential  vanishes, $V|_{min}=0$, then at this point
the kinetic energy becomes $\dot T^2|_{min}=1$. Therefore,
if we want to have a canonically normalized field which
includes the late time behavior of $T$, specially important for dark energy
considerations, then the field transformation should
necessarily be valid for large values of $\dot T$.

\section{Mass}\la{secM}

The mass of a particle enters in different phenomenological
processes such as particle decay   or reheating. Therefore, it
is important to determine what is the mass of the tachyon field $T$.
 Naively,  the mass of  a scalar field
is given by the second derivative of the potential $V(\phi)$ w.r.t. the field,
e.g. $m^2=dV^2/d\phi^2$. However, this statement is only valid if the scalar field
has a canonical kinetic term. Of course, this is not the case for the tachyon field $T$.
Even if we consider the effective potential $F$, c.f. eq.(\ref{F}), the mass
of $T$ is neither given by $d^2V/dT^2$ nor by $d^2F/dT^2=V_{TT}/V-V_T^2/V^2$.

The mass of a particle is simply given by the pole of the propagator
\ci{weinberg}. So in order to determine what the mass of the tachyon
field is, we need to expand the Lagrangian $L(\phi,\pu\phi)$ up
 to second
order in the field perturbations $\phi(t,x)\simeq \phi_o(t)+\delta\phi(t,x)$. For quite arbitrary
potentials and kinetic terms with $L(\phi,\pu\phi\pU\phi)$  ,  the second
order term can be put in the form
\bea\label{dL1}
\int dx^4 \delta L&=& \int dx^4  [A(t) \delta\dot \phi^2 + B(t) \delta\phi\delta\dot\phi +\\ \non
&+ & C(t) \delta\phi'^2 +D(t) \delta\phi^2] ,
\eea
with $\delta\phi'=\pp(\delta\phi)/\pp x$ and
\bea\la{A0}
A=\frac{1}{2}\frac{\pp^2L}{\pp\dot\phi^2},&& B=\frac{\pp^2L}{\pp\dot\phi\pp\phi},\\
C=\frac{1}{2}\frac{\pp^2L}{\pp\phi'^2},&&D=\frac{1}{2}\frac{\pp^2L}{\pp\phi^2} ,
\eea
with all other second derivative terms being zero.
After integrating by parts and dropping surface terms we get
\be\la{dL2}
\int dx^4 \delta L=-\int dx^4 A\;\delta\phi \le( \pp_t^2
+\frac{C}{A} \pp_x^2 +m^2 \ri)\delta\phi ,
\ee
with the mass of the scalar field given by
\be\la{m}
m^2=-\frac{1}{A}\le(D+\frac{\ddot A}{2}-\frac{\dot B}{2}\ri).
\ee
Clearly from eq.(\ref{dL2}), $m^2$ represents the pole of the propagator $(\pp_t^2+\frac{C}{A}
\pp_x^2 + m^2  )\Delta_F(x)=-\delta^4(x)$.

In the case of a canonical normalized scalar field with Lagrangian
$L=(\pu \varphi)^2/2 - W(\varphi)$ the coefficients of
eq.(\ref{dL1}) are
\bea\la{A1} A=\frac{1}{2},&&B =0,\\ \non
 C=-\frac{1}{2}, &&D=-\frac{1}{2}\frac{d^2W}{d\varphi^2}.
 \eea
The term $A=1/2$ defines a canonically normalized field and
the mass is given by the usual expression $m^2=-D/A=d^2W/d\varphi^2$.

In the case of the tachyon field $T$ given in eq.(\ref{L}), one has
\bea\la{A}
A=\frac{\rho^3}{2V^2}, &&
B=\frac{\rho V_T\dot T}{V}, \\ \non
C=-\frac{\rho}{2},&&
D=-\frac{V V_{TT}}{2\rho} ,
\eea
which gives a mass term
\be\la{mT}
m^2(T)=\frac{V_{TT}}{V}-\frac{V_T^2}{V^2}\le(3-\frac{V^2}{\rho^2}\ri) ,
\ee
were we have used eq.(\ref{m}). As mentioned above, the pole
of the propagator, which gives the mass of the tachyon,
does not coincide  with $d^2V/dT^2$ nor with $d^2F/dT^2=V_{TT}/V-V_T^2/V^2$.

The term proportional to the space dimensions  $\pp_x^2$ is given by
$C/A=-V^2/\rho^2$ giving a  Carrollian type metric (for further
reading see \ci{Gibbons}). The phase velocity of the perturbations
is given by $v^2_{ph}=-C/A=V^2/\rho^2$ and it vanishes at $V=0$.

\section{Normalized Scalar Field}\la{secN}

We are now interested in finding a field transformation which gives
a canonical tachyon field valid for arbitrary values of $\dot T$. A
scalar field with canonical kinetic terms in the limit of small
kinetic term ($|\dot T|\ll 1$) has been studied in the literature,
giving a Lagrangian $L\simeq  V(T)(1-\dot T^2/2)$ with $A=V/2$ as
defined in eq.(\ref{A0}), and suggesting the transformation \ci{slowroll}
\be\la{dp2}
\dot \varphi= \sqrt{V}\;\dot T.
\ee
This transformation
gives a Lagrangian $L\simeq \dot\varphi^2/2-V$.
 However, clearly the field
$\varphi$ is  canonically normalized only in the limit of small kinetic terms
and as discussed above the most interesting region is at late times where $V$
approaches zero and $\dot T\simeq 1$.

Instead,  let us expand eq.(\ref{L}) around an arbitrary value of $
T_o(t)$ and $\dot T_o(t)$, i.e.
$T(t)=T_o(t)+\delta T(t,x)$,  and the  second order term in
$\delta\dot T$ is given by
\be\la{dL}
\delta L\simeq
\fr{V(T_o)}{2(1-\dot T_o^2)^{3/2}}\;\delta\dot T^2 ,
 \ee
 which is just
the term $A$  of eq.(\ref{A}) since the energy density
$\rho=V(T_o)/\sqrt{1-\dot T_o^2 }$ is constant. From eq.(\ref{dL}) we
suggest the new  tachyon transformation
\be\la{dp} \dot\phi=
\sqrt{\fr{V(T_o)}{(1-\dot T_o^2)^{3/2}} } \;\;\dot T
=\fr{\rho^{3/2}}{V} \;\;\dot T ,
\ee
to obtain a canonical normalized
tachyon field $\phi$. The transformation in eq.(\ref{dp}) reduces to
that of eq.(\ref{dp2}) when $|\dot T| \ll 1$ which implies that
$\rho\simeq V$. However, in our case the transformation in
eq.(\ref{dp}) is not only valid for small $\dot T$ but it is also
valid for arbitrary values of $\dot T$.

Integrating eq.(\ref{dp}), using  the fact that $\rho(t)$ is constant, we get
\be\la{phi}
\phi= \int \fr{\rho^{3/2}\dot T}{V}\,dt=\rho^{3/2} \int \fr{\dot T}{V}\,dt\,=\,\rho^{3/2} \int \fr{dT}{V}.
\ee

The equation of motion for $\phi$ can be easily obtained from eq.(\ref{eqM})
using eq.(\ref{dp}) (with $\rho$ constant) giving
\bea\la{eqMp}
\ddot\phi + \frac{\rho^3}{V^3}\,\frac{dV}{d\phi}&=&0 ,\\
\ddot\phi + U_\phi &=&0 ,
\eea
with the effective potential $U(\phi)$ defined by $U_\phi\equiv dU/d\phi=(\rho^3/V^3)dV/d\phi$.
For $\rho$ constant we can integrate $U$ and we get an effective potential
\be\la{U}
U=U_o -\frac{\rho^3}{2V^2} ,
\ee
with $U_o=3\rho/2$. The energy density in eq.(\ref{rho}) can be expressed in terms
of $\phi$ and the effective potential $U$ as,
\be\la{rp}
\rho=\frac{V}{\sqrt{1-\dot T^2}}=\frac{V}{\sqrt{1-(V^2/\rho^3)\dot \phi^2}}
=\frac{1}{2}\dot\phi^2+U(\phi).
\ee
The last equality  in eq.(\ref{rp}) can be easily derived
by extracting $\dot\phi^2$ from $\rho^2= V^2/(1-V^2\dot \phi^2/\rho^3)$
and comparing it with $\dot\phi^2=2(\rho-U)$.
It is clear from the last equality  in eq.(\ref{rp})
that the energy density $\rho$
in terms of $\phi$ and $U$ is that of a canonically normalized scalar field which can
be derived from a Lagrangian
\be\la{Lp}
L=\fr{1}{2}\dot\phi^2-U(\phi) ,
\ee
and with an equation of motion given by eq.(\ref{eqMp}). The mass of the
$\phi$ is given by
\be\la{mp}
M^2(\phi)=\frac{d^2U}{d\phi^2}=\frac{\rho^3}{V^2}\le(\frac{V_{\phi\phi}}{V}-3\frac{V_\phi^2}{V^2}\ri) ,
\ee
where $V_\phi\equiv dV/d\phi$. In terms of the field $T$ using eq.(\ref{dp}) we can
express $U_\phi$ and $U_{\phi\phi}$ as
\bea\la{ut}
U_\phi&=&\frac{\rho^3}{V^3}\frac{dV}{d\phi}=\frac{\rho^3}{V^3}\frac{dT}{d\phi}
\frac{dV}{dT}=\frac{\rho^{3/2}}{V^2}\frac{dV}{dT} ,\\
M^2&=&\frac{d^2U}{d\phi^2}=\frac{\rho^3}{V^2}\le(\frac{V_{\phi\phi}}{V}-3\frac{V_\phi^2}{V^2}\ri)
= \le(\frac{V_{TT}}{V}-2\frac{V_T^2}{V^2}\ri).
\la{utt}\eea

If we substitute eq.(\ref{dp}) into eq.(\ref{L}) we get a
Lagrangian $L=-V(\phi)\sqrt{1-V(\phi)^2\dot\phi^2/\rho^3}$, with $\rho$
constant, and if we expand to second order
in perturbations of $\phi$ around an arbitrary value $\phi_o(t)$, the coefficients
$A,B,C,D$, as defined in eq.(\ref{dL1}), are  now given by
\bea\la{A2}
A=\frac{1}{2}, && B=\frac{\dot\phi V_\phi}{V}\le(3-\frac{2V^2\dot\phi^2}{\rho^3}\ri) ,\\ \non
C=-\frac{V^2}{2\rho^2}, && D=-\frac{V_{\phi\phi}\rho}{2V}(1-2V^2\dot\phi^2)+
\frac{V_\phi^2\dot\phi^2}{2V^2}(3-2V^2\dot\phi^2).
\eea
In this case we have $A=1/2$ and $C/A=-V^2/\rho^2$ showing that the field
$\phi$ has canonical normalized kinetic
term (in a Carrollian metric).
The mass of $\phi$ given as the pole of the propagator using eqs.(\ref{m}) and (\ref{A2})
gives a mass which is equal to that of eq.(\ref{mp}).

We have shown that to first and second order
 the homogenous part of the Lagrangian $L=-V(\phi)\sqrt{1-V^2\dot\phi^2/\rho^3}$
 and $L=\fr{1}{2}\dot\phi^2-U(\phi)$
 are equivalent.

\section{General Analysis}\la{secG}

Let us now study some general properties  of the tachyon field $\phi(T)$.
Without loss of generality we take the origin at $T=0$ and we
assume $V_i=V(T=0)\neq 0 $. If the field $T$ is a tachyon at the origin
then $V_T(T=0)=0$ and $V_{TT}(T=0)>0$. We can see that the tachyonic
property of $T$ is transmitted to $\phi$ and its potential $U(\phi)$, since
at the origin $\phi=0$ and from eqs.(\ref{U}), (\ref{ut}) and (\ref{utt}) we have
$U(\phi=0)=\rho=V_i$ and $U_\phi(0)=V_T/V_i^{1/2}|_{min}=0$ and
$U_{\phi\phi}(0)=V_{TT}/V_i|_{min} > 0$.

The potential $U=U_o-\rho^3/2V^2$  diverges
and is unbounded from below at $V\rightarrow 0$.
In terms of the potential $U$
and using eq.(\ref{dp}) the value of $\dot\phi$    is
\be
\dot\phi^2=2(\rho -U)=\rho\le(\frac{\rho^2}{V^2}-1\ri).
\ee
At the origin one has $V(T_i)=V_i=\rho$  and $\dot\phi=0$  while
at the minimum $V(T)=0$ one has $|\dot\phi|=\infty$.
Even though
the potential is unbounded from below the total energy density $\rho$ remains
constant.  The
conclusion is that at the minimum the quantities $U(\phi)$ and the
field $\phi$, $\dot\phi$  tend
to an infinite value, showing that $\phi$ has
a runaway behavior and does not oscillate at the minimum $V=0$. However, there
is no stable configuration. This could be interpreted as the decay of the
unstable brane-antibrane system represented by  the tachyonic potential.

\subsection{Examples}

In this  section we will analyze a few interesting examples.
First we will study the potentials $V=V_i\, Exp\,[-T^2/2]$ \ci{tachsen}
and $V=V_i/\cosh [T]$ \ci{cosh}. Both of them have been conjectured
to represent the tachyon field for a Dp-brane system
in superstrings. Later we will study  a cubic potential
which is unbounded from below for negative $T$ and
has been related to Dp-brane system
in bosonic strings.

\subsubsection{Potential $V=V_i\, Exp\,[-T^2/2]$}

The potential $V=V_i\, Exp\,[-T^2/2]$ has been widely studied in the
literature \ci{DB2},\ci{tachsen}. It has a maximum at $T=0$ while it goes to zero at
large $T$ with vanishing potential.

\begin{figure}[tbp]
      \includegraphics[width=6cm]{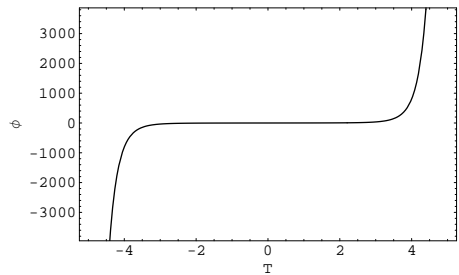}
  \caption{ We show the normalized tachyon field $\phi$ as a function of $T$
  for $V=V_iExp[-T^2/2]$.}
 \label{1c}
\end{figure}

\begin{figure}[tbp]
      \includegraphics[width=6cm]{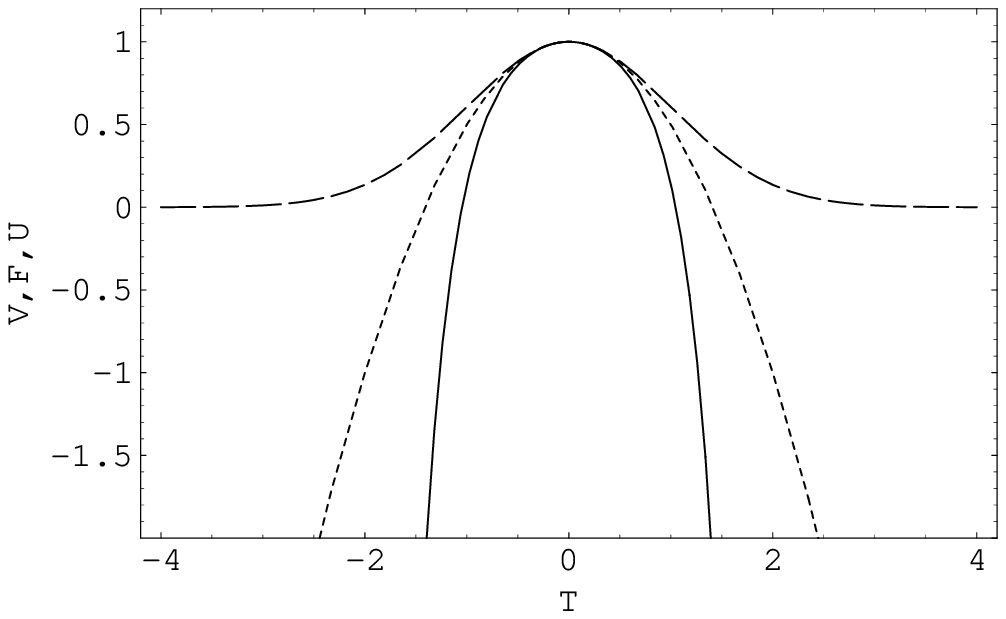}
  \caption{ We  show the potentials $U,V,F$ as a function of $T$
  (solid, dashed and dotted lines respectively) for $V=V_iExp[-T^2/2]$.}
 \label{1a}
\end{figure}

The normalized tachyon field $\phi$ in terms of $T$, as
given by eq.(\ref{phi}), is
\be
\phi= \sqrt{\fr{\pi}{2}}\,E_{rfi}\le[\frac{T}{\sqrt{2}}\ri] ,
\ee
and can be seen in figure (\ref{1c}). Notice that at
$T\rightarrow \pm \infty$ one has the limit $\phi\rightarrow \pm \infty$.
The effective potential $U(\phi)$ is unbounded from below for
$V(T)\rightarrow 0$ and we show in fig.(\ref{1a}) the behavior of $U$, $V$
and $F=Log[V] + c$, with $c$ a constant such that $F(T_i)=V(T_i)$
given by a solid, dashed and dotted lines respectively. The
effective potentials $ F$ is also unbounded from below for $V(T)\rightarrow 0$.

\begin{figure}[tbp]
      \includegraphics[width=6cm]{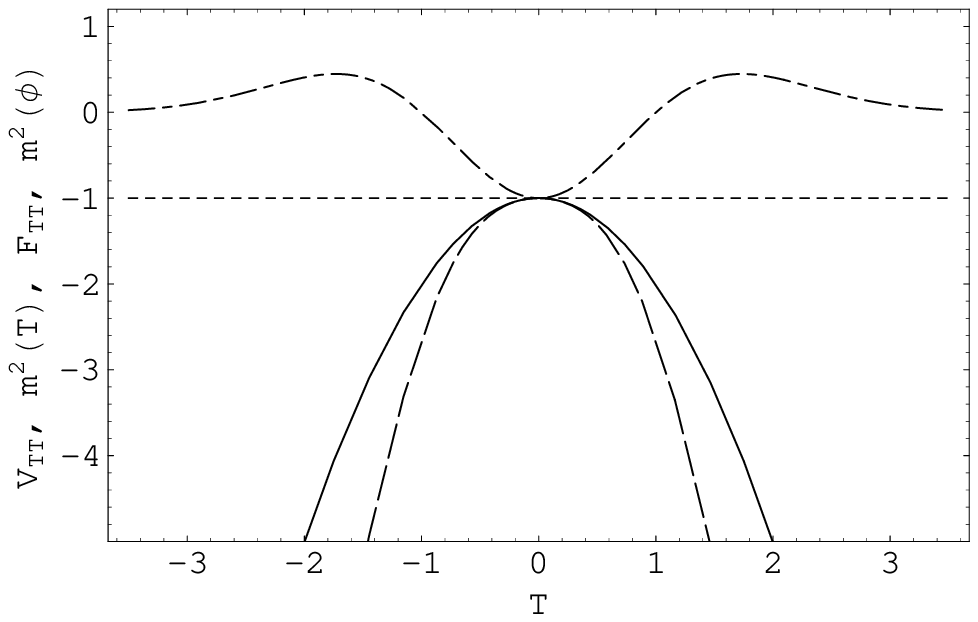}
  \caption{ We  show the mass $M^2 , \,m^2 $
  and the second derivatives $V_{TT}, F_{TT}$ as a function of $T$
 (solid, dashed, dash-dotted and dotted, respectively) for $V=V_iExp[-T^2/2]$.}
 \label{1b}
\end{figure}

The mass of the normalized tachyon field $\phi$ given by eq.(\ref{utt}),
 the mass of $T$ given by eq.(\ref{mT}), and the second
derivative of $V$ and $F$ are then
\bea
M^2&=&-(T^2+1),\; m^2=T^2(e^{-T^2}-2)-1 ,\\
 V_{TT}&=&e^{-T^2/2}(T^2-1),\;\;F_{TT}=-1 ,
 \eea
and are shown in fig.(\ref{1b}) (solid, dashed, dash-dotted and dotted, respectively).
The mass $M^2, m^2, F_{TT}$  are negative for all values of $T$ and $M^2, m^2$ diverge
at $T\rightarrow\pm \infty$ or equivalently at $V(T)\rightarrow 0$. The infinite
mass implies that the field is no longer a dynamical field at these points. On the other hand,
the quantity $V_{TT}$ is negative at the origin and tends to zero from above
at $T=\pm \infty$.


\subsubsection{Potential $V=V_i/\cosh [T]$}

Let us now  study the potential $V=V_i/Cosh[T]$ \ci{cosh}, which  has
 been also widely analyzed in the literature. This potential
 has a maximum at the origin $T=0$, is never negative, and goes to zero
 at large $T$.

\begin{figure}[tbp]
      \includegraphics[width=6cm]{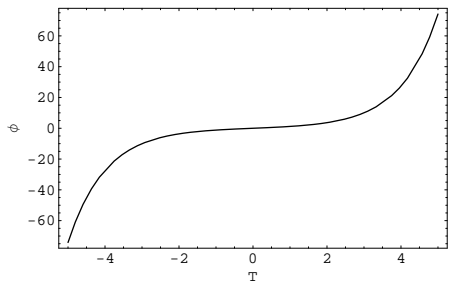}
  \caption{ We show the normalized tachyon field $\phi$ as a function of $T$
  for a potential $V=V_i/\cosh[T]$.}
 \label{2c}
\end{figure}

\begin{figure}[tbp]
      \includegraphics[width=6cm]{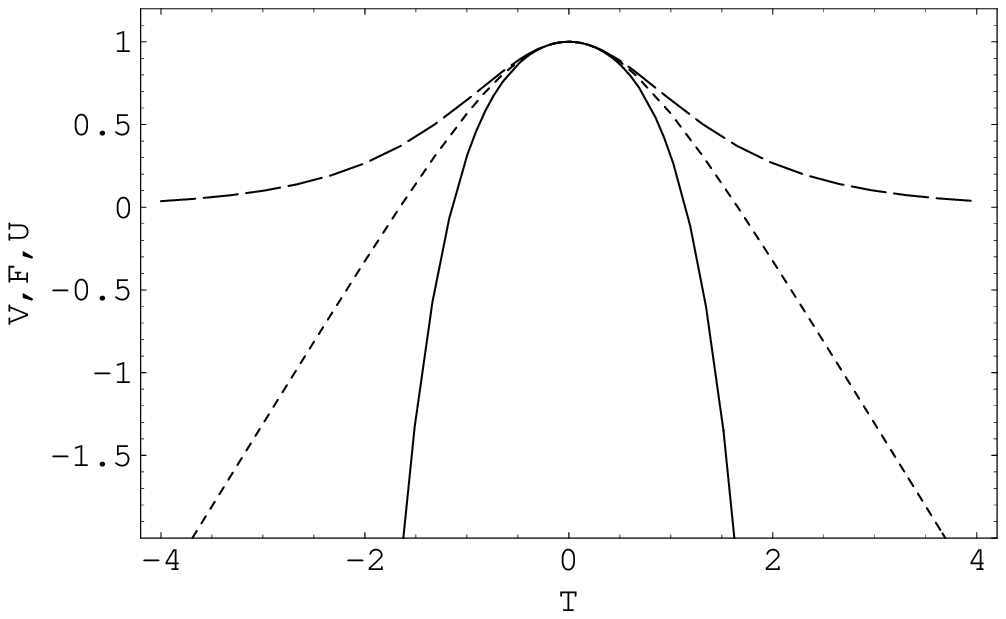}
  \caption{ We  show the potentials $U,V,F$ as a function of $T$
  (solid, dashed and dotted lines respectively)  for  $V=V_i/\cosh[T]$.}
 \label{2a}
\end{figure}

Using eq.(\ref{phi}) the normalized tachyon field $\phi$  in terms of $T$
   is
\be
\phi= \sinh [T] ,
\ee
and can be seen in figure (\ref{2c}). For
$T\rightarrow \pm \infty$ one has the limit $\phi\rightarrow \pm \infty$,
as expected.
The effective potentials $U(\phi)$ and $F(T)$ are unbounded from below for
$V(T)\rightarrow 0$ while $V(T) \geq 0$ for all $T$. In fig.(\ref{2a})
we show the behavior of $U$, $V$
and $F $ given by the solid, dashed and dotted lines respectively. The
behavior of the normalized scalar field $\phi$ and the potentials $U, F$ are very similar
in the two examples considered, namely  $V=V_iExp[-T^2/2]$ and $V=V_i/\cosh[T]$.

\begin{figure}[tbp]
      \includegraphics[width=6cm]{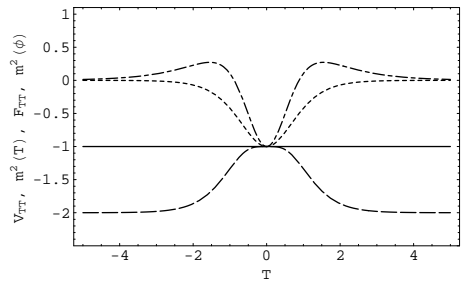}
  \caption{ We  show the mass $M^2 , \,m^2 $
  and the second derivatives $V_{TT}, F_{TT}$ as a function of $T$
 (solid, dashed, dash-dotted and dotted, respectively) for  $V=V_i/\cosh[T]$.}
 \label{2b}
\end{figure}

The mass of the normalized tachyon field $\phi$ given by eq.(\ref{utt}),
 the mass of $T$ given by eq.(\ref{mT}), and the second
derivative of $V$ and $F$ are in this case
\bea
M^2&=& -1,\;\;   m^2=-\frac{1}{4}\frac{\le(\cosh [4T]+3\ri)}{\cosh [T]^{4}} ,  \\
 V_{TT}&=&\frac{1}{2}\frac{\le(\cosh [2T]-3\ri)}{\cosh [T]^{3}},\;\;   F_{TT}=-\frac{1}{\cosh [T]^{2}} ,
 \eea
and are shown in fig.(\ref{2b}) (solid, dashed, dash-dotted and dotted, respectively).
The mass $M^2$ is negative and constant, while $m^2, F_{TT}$
are negative for all values of $T$   and approach zero from below
for  $T\rightarrow\infty$.     The
quantity $V_{TT}$ is negative at the origin and tends to zero from
above at $T=\pm\infty$.

Contrary to the case $V=V_iExp[-T^2/2]$ for  the potential $V=V_i/\cosh[T]$
 the mass of $\phi$ and $T$, i.e. $M^2$ and $ m^2$,  do not diverge even though the effective
 potentials $U,\, F$ diverge at large $T$.


\subsubsection{Potential $V=T^3+T^2-4/27$}

\begin{figure}[tbp]
      \includegraphics[width=6cm]{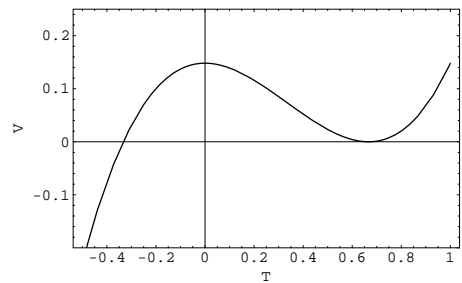}
  \caption{ We show the potential  $V=T^3+T^2-4/27$ as a function of $T$.}
 \label{3d}
\end{figure}

We will now  analyze the potential $V=T^3+T^2-4/27$.
This kind of potential has been suggested \ci{tachsen} to parameterize
a tachyon field  of a unstable bosonic Dp-brane systems.
The potential has a maximum at the origin $T=0$,  a minimum
at $T=2/3$,  is unbounded from below for negative $T$ and
 $V(T)$ vanishes at $T=2/3,-1/3$  as seen from fig.(\ref{3d}).

\begin{figure}[tbp]
      \includegraphics[width=6cm]{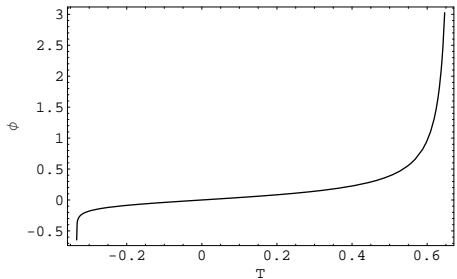}
  \caption{ We show the normalized tachyon field $\phi$ as a function of $T$.}
 \label{3c}
\end{figure}

\begin{figure}[tbp]
      \includegraphics[width=6cm]{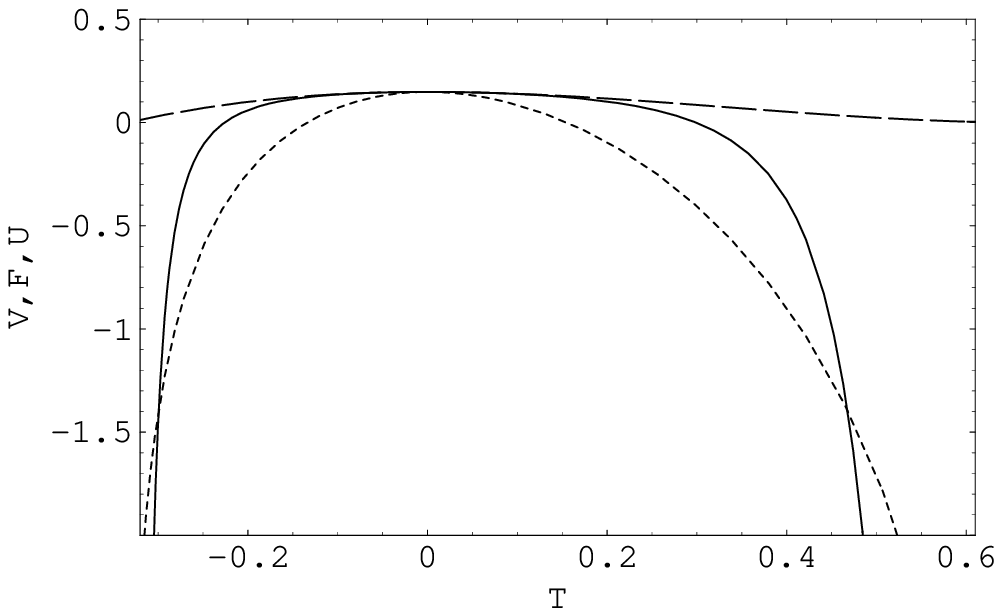}
  \caption{ We  show the potentials $U,V,F$ as a function of $T$
  (solid, dashed and dotted lines respectively).}
 \label{3a}
\end{figure}

The normalized tachyon field $\phi$ in terms of $T$, as
given by eq.(\ref{phi}), is
\be
\phi=\frac{8}{81\sqrt{3}}\le(\frac{9T}{2(2-3T)}+Log\le[\frac{2(1+3T)}{2-3T}\ri]\ri) ,
\ee
and can be seen in figure (\ref{3c}). It is clear that $\phi$
has a pole at $T=2/3 $ and $T= -1/3$ where $|\phi|\rightarrow  \infty$.
So, even though the potential $V(T)$ is defined for all values of
$T$, the normalized tachyon $\phi$  is only  defined in the interval $-1/3\leq T \leq 2/3$
which implies that the field $\phi$ does not know of the unboundedness
of $V(T)$ for negative $T$. However, once again
the effective potential $U(\phi)$ and $F$  are unbounded from below for
$V(T)\rightarrow 0$ and we show in fig.(\ref{3a}) the behavior of $U$, $V$
and $F $ given by a solid, dashed and dotted lines, respectively.

\begin{figure}[tbp]
      \includegraphics[width=6cm]{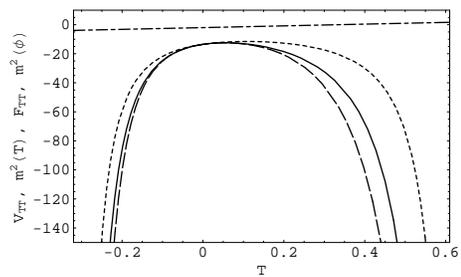}
  \caption{ We  show the mass $M^2(\phi), \,m^2(T)$
  and the second derivatives $V_{TT}, F_{TT}$ as a function of $T$
 (solid, dashed, dash-dotted and dotted, respectively).}
 \label{3b}
\end{figure}

The mass of the normalized tachyon field $\phi$ given by eq.(\ref{utt}),
 the mass of $T$ given by eq.(\ref{mT}), and the second
derivative of $V$ and $F$ are then
\bea
M^2&=& -\frac{54(1+18T^2)}{(2-3T)^2(1+3T)^2},\\
  m^2&=&\frac{27[-32-576\,T^2+5832\,T^4(T-1)]}{16(2-3T)^2(1+3T)^2},\\
 &+&\frac{27\times19683\,T^6(1-T)^2}{16(2-3T)^2(1+3T)^2},\\
 V_{TT}&=&6T-2,\\
   F_{TT}&=&-\frac{27(2+9T^2)}{(2-3T)^2(1+3T)^2} ,
 \eea
and are shown in fig.(\ref{3b}) (solid, dashed, dash-dotted and dotted, respectively).
The mass $M^2, m^2$ and $F_{TT}$  are negative for all values of $T$ and have a pole at
$T=2/3, -1/3$ or equivalently $V(T)\rightarrow 0$. The infinite
mass implies that the field is no longer dynamical at these points. The
quantity $V_{TT}$ is only positive for $T>1/3$.

\section{Late time behavior}\la{secL}

We will analyze  the late time
behavior of the normalized tachyon field
in a BI Lagrangian. We are also interested in
determining if we can obtain an accelerating
universe at late times.
In order to do this, we will
 use the L'H\^{o}pital rule to derive
the slow roll conditions which will given by
exact relationships. Tachyonic  slow
roll conditions has been studied in \ci{slowroll}.

\subsection{Slow roll conditions and L'H\^{o}pital rule}

\subsubsection{Canonical scalar field $\varphi$}

As a matter of presentation purpose let us first begin with  a
canonically normalized field
and then we will study the case of a BI type lagrangian.
In this first case the energy density is  given by $\rho=E_k+V$
and the pressure by $p=E_k-V$ with $E_k=\dot\varphi^2/2$ the kinetic energy and
$V(\varphi)$ the scalar potential. A general analysis can be found in \ci{ax}. The evolution of $\rho$ and acceleration
of the scale factor $a(t)$ of the universe are given in terms of $\rho,p$ by
\bea\la{sr}
\dot\rho&=&-3H(\rho+p)=-6H\rho\le(1-X\ri),\\
\frac{\ddot a}{a}&=&-\frac{1}{6}\le(\rho+3p\ri)=- \rho\le(\frac{2}{3}-X\ri) ,
\la{sa}\eea
with   $H\equiv \dot a/a$. We have defined the quantity $X\equiv V/\rho$
and we expressed $E_k=\rho-V=\rho(1-X)$ in the last part of eqs.(\ref{sr}) and (\ref{sa}).
The quantity $X$ is  constrained  by the values $1\geq X\geq 0$.
From eq.(\ref{sa}) we see that if $X $ is larger than
$2/3$ then we have an accelerating universe. The equation of state
parameter $w$ is given by $w=p/\rho=1-2X$ and if $X$ is constant then
we have a scaling solution with $\rho=\rho_o(a/a_o)^{-6(1-X)}$.

We will consider only potentials that vanish at the minimum $V_{min}=0$ (if $V_{min}>0$
then it is easy to show that the dynamics always leads to inflation and for
$V_{min}<0$ we have a non-inflationary universe and in fact a big crunch \ci{ax.neg}).
For $V=0$ with $\rho\neq 0$, i.e. $X=0$, we see from eq.(\ref{sa}) that $\varphi$
does not lead to an inflationary universe. However, from eq.(\ref{sr})
we have in this case a negative $\dot\rho$
and therefore $\rho$ decreases until it reaches $\rho=0$. This arguments shows
that the late time behavior is the expected one $V\rightarrow 0, \,\rho\rightarrow 0$
but the limit $X=V/\rho$ is model dependent. Since the limits of $V$ and $\rho$ are
zero we can use the L'H\^{o}pital rule to determine the asymptotic (late time) value,
\be\la{H1}
 \lim \le(X=\frac{V}{\rho }\ri)=\lim\le( \frac{\dot V}{ \dot\rho  }\ri) ,
 \ee
 and since the limit $  \dot\rho=0$, if $\dot V$ also vanishes  then
 \be
\lim X = \lim\le( \frac{\ddot V}{ \ddot\rho  }\ri).
\la{H2}\ee
Now, taking $H^2 =\rho/3$,  the equation of motion $\ddot\varphi=-V'-3H\dot\varphi$
and using eq.(\ref{sr}) the
 equations (\ref{H1}) and (\ref{H2}) give the constraints
\bea\la{slr1}
\frac{V'}{V}&=&\sqrt{6(1-X)}< \sqrt{2},\\
\frac{V''}{V}&=&6(1-X)< 2 ,
\la{slr1a}\eea
where we  have used eq.(\ref{slr1}) to write  the second derivative of the potential as
 $\ddot V(\varphi)=V''\dot\varphi^2+V'\ddot\varphi=
2(1-X)\rho(V''+3V(1-X))$ to derive eq.(\ref{slr1a}).

Equations (\ref{slr1}) and (\ref{slr1a}) are the slow roll conditions at late time (i.e. for vanishing
$V,\rho$) and are exact equations in this limit. An inflationary
universe must satisfied the condition $X>2/3$ which gives the last inequalities
in eqs.({\ref{slr1}). We have determined that exact value of $V'/V, V''/V$ at which
the universe no longer inflates at late time.  It is easy to see that
the limit is satisfied for a potential of the form $V=V_o\,exp[6(1-X)\phi]$
and the exponent must be smaller than two for $V$ to inflate. This potential
gives the scaling solution (i.e. $X=cte$) and $\rho \propto(a/a_o)^{-6(1-X)}$.

A late time inflation, i.e. dark energy, requires $X>2/3$ and in the limit $X\rightarrow 1$
from eqs.(\ref{slr1}) one has  $V'/V\rightarrow 0, V''/V\rightarrow 0$. For a
potential $V=\varphi^\alpha$ a  period of late time inflation is simply obtained for
a negative  $\alpha$.

\subsubsection{Tachyon field $T$}

Now, let us repeat the analysis for the tachyon field with Lagrangian (\ref{L})
in a FRW metric. The equation of motion for $T$ is $\ddot T/(1-\dot T^2)+3H\dot T+V_T/V=0$
and the energy density and pressure are given by eqs.(\ref{rho}). The evolution
of $\rho$ and the acceleration of the universe are now given by
\bea\la{sr2}
\dot\rho&=&-3H(\rho+p)=-3H\rho\le(1-X^2\ri),\\
\frac{\ddot a}{a}&=&-\frac{1}{6}\le(\rho+3p\ri)=- \frac{1}{2}\rho\le(\frac{1}{3}-X^2\ri) ,
\la{sa2}\eea
with $1\geq X=V/\rho\geq 0$.
In this case an inflationary universe requires $X^2>1/3$.
It is clear that the r.h.s. of eq.(\ref{sr2}) is non positive
which implies that $\rho $ decreases or it remains constant in time.
The constant value of $\rho$ takes place either for $\rho=0$
or at $X^2=V^2/\rho^2=1$. As in the canonical case,
if $V(T)=0$ with $\rho\neq 0$ then $X=0$ and from
eq.(\ref{sa2}) it is easy to see that the tachyon field
does not inflate the universe. However, in the limit
$V\rightarrow 0$ with $X\neq 1$ the energy
density decreases and $ \rho\rightarrow 0$ at late times.

Since $V$ and $\rho$ tend to zero then we can use the L'H\^{o}pital rule to
determine the asymptotic behavior of $X$ and using the equation
of motion of $T$ the eqs.(\ref{H1}) and (\ref{H2})
give  the constraints
\bea\la{sl2}
|\frac{V_T}{V^{3/2}}&=& - \sqrt{3\le(\frac{1}{X}-X\ri)}|< \sqrt{2\sqrt{3}}, \\
\frac{V_{TT}}{V^{2}}&=&\frac{9}{2}\le(\frac{1}{X}-X\ri) < 3\sqrt{3} .
\la{s12b}\eea
Eqs.(\ref{sl2}) and (\ref{s12b})
 give the exact late time limit for tachyonic potentials and
the inequality must hold for the universe to inflate (i.e. $X^2>1/3$).
A scaling solution requires $X=cte$ and from eqs.(\ref{sl2}) we obtain
the potential $V=V_o/T^2$. This potential has been widely studied \ci{T2}
and here we have derived it through a very simple and powerful tool,
the L'H\^{o}pital rule. Our analysis reproduces the result for a  scaling solution
given in \ci{tachCopeland}, namely $\lambda\equiv -V_T/V^{3/2}$ constant and $ \Gamma\equiv VV_{TT}/V_T^2=3/2$.
Using the potential $V=V_o/T^2$ in eqs.(\ref{sl2})
the constant $V_o$ is given in terms of $X$ by $V_o=4X/3(1-X^2)$
and an inflationary universe requires $V_o\geq 2/\sqrt{3}$ for$\,X^2\geq 1/3$. Contrary to
a canonically normalized fields were an inverse power potential
always leads to inflation at late times in this case the constant
term determines whether the universe inflates or not.

If $X=0$ then the r.h.s. of eqs.(\ref{sl2}) and (\ref{s12b})
is infinite and taking the ansatz $V=T^\alpha$, giving
$V_T/V^{3/2}=T^{-(2+\alpha)/2}$ and $ V_{TT}/V^{2}=T^{-(2+\alpha)}$,
eqs.(\ref{sl2}) and (\ref{s12b})  are satisfied for $\alpha > 0$ with $
V(T\rightarrow 0)\rightarrow 0$ or for $\alpha <-2$ in the limit $
V(T\rightarrow \infty)\rightarrow 0$.  On the other hand,
if $X=1$ then the r.h.s. of eqs.(\ref{sl2}) and (\ref{s12b})
vanish which implies  $0\geq \alpha >-2$ for $
V(T\rightarrow \infty)\rightarrow 0$. We have, therefore, shown
that for inverse power potentials only the models with
$0\geq  \alpha \geq -2$ lead to a late time inflation, i.e.
dark energy.

\subsection{Asymptotic Limit for $\phi$}

Now, we would like to study how the normalized tachyon field $\phi$ defined
by eq.(\ref{dp}) behaves at late time.
We will assume that the origin is at $T=0$
and the    potential $V(T)$ vanishes
at $T=T_m$, i.e. $V(T_m)=0$, where $T_m$ can take
either an  infinite or a finite value. The
potential may have a minimum at $T_m$ but it is not
relevant if $V(T_m)$ is a minimum or not.

Let us first begin the study in a Minkowski metric. In this
case we have a  constant  $\rho$ and from eq.(\ref{dp}) it is easy
to see that  in the limit $V(T\rightarrow T_m)\rightarrow 0$  one has
\be
\phi=\int \rho^{3/2} \frac{dT}{V(T)}=\rho^{3/2}\int
 \frac{dT}{V(T)}\rightarrow\infty ,
\ee
since the integrand diverges. This implies that the  normalized scalar
field $ \phi$
is constraint between the zeros of the potential $V(T)$ and never reaches the
region of $T>T_m>0$ (or $T<T_m$ for negative $T_m$)  and $\phi$ always reaches an infinite
value at $V(T)=0$, i.e. the field $\phi$   is only defined in the interval
$T_{m1}\leq T\leq T_{m2}$, with $V(T_{m1})=V(T_{m2})=0$,
and  $\phi$ takes  the whole range of values   $-\infty\leq\phi\leq\infty$.
This result contrast with the one obtained by using the transformation in eq.(\ref{dp2})
where finite values for $\varphi$ are obtained in the limit $V\rightarrow 0$.
 However, we would like to point
out again that the transformation (\ref{dp2}) gives a canonical scalar field
only for small values of $\dot T$ and not for values of $\dot T\simeq 1$ where
$V$ approaches zero.

The constraint on the interval on the
tachyon field $T_{m1}\leq T\leq T_{m2}$, with $V(T_{m1})=V(T_{m2})=0$
is quite important  for
models with finite values $T_m$ as in a cubic potential $V=T^3+T^2-c,(c>0)$,
where it has a local minimum for $T>0$ and is unbounded from below for $T<0$.
This kind of potentials, obtained from Dp-brane in bosonic string theory,
have not
been thoroughly analyzed because for $T<0$ the potential $V(T)$
it is argued that the potential is unbounded
from below and for the  minimum  at positive $T$ the system is unstable
(a large growth of perturbations when $T$ oscillates around $T_m>0$).
However, the physical field $\phi$ does not reach the regions outside
the zeros of the potential $V(T)$ and therefore does not oscillate
around the minimum with  $T_m>0$ nor does it feel the unboundedness
of the potential for negative $T$. The fact that $\phi$ does not
oscillate around the minimum is what one would expect from
a string motivated tachyon \ci{tachsen}.

\subsubsection{Asymptotic Limit for $\phi$ in FRW}

In the case of a FRW metric, the energy density is no longer constant
and we cannot take $\rho$ out of the integral in eq.(\ref{phi}). However,
we can still easily investigate the asymptotic value of $\phi$
for $V(T)\rightarrow 0$ using the L'H\^{o}pital rule, as in the previous section.
The relevant quantity is the integrand $Y\equiv \rho^{3/2}/V(T)$
of eq.(\ref{phi}).
Similar to eqs.(\ref{H1}) and (\ref{H2}) we have in this case
\be\la{Y}
\lim \le(Y=\frac{\rho^{3/2}}{V}\ri)=\lim\le( \frac{(3/2)\dot\rho \rho^{1/2}}{\dot V}\ri),
\ee
with $\dot V= V_T \dot T$.
Using eqs.(\ref{sr2}) and (\ref{sl2}),    equation (\ref{Y}) becomes
\be\la{Y1}
\lim Y =\lim \frac{3}{2}\; Y,
\ee
which is only satisfied if $Y=\rho^{3/2}/V=0$ or $Y=\infty$.
Since $Y=\sqrt{\rho}/X=\sqrt{V}/X^{3/2}$ and $0\geq X=V/\rho \geq 1$ a value
of $X\neq 0$ gives $Y\rightarrow 0$. This is
the case for any model that leads to late time inflation
since $X$ must be larger than $2/3$. In this case,
the integrand in eq.(\ref{phi}) vanishes at late times
and a finite value of $\phi$ is obtained
as long as $\int Y dT\simeq (\int \sqrt{V} dT)/X^{3/2}$
is finite in the limit of vanishing $V$ (we have taken $X$ out of the integral
since in the late time limit it  approaches
a constant value).  This will
happen for models
with a vanishing potential $V(T)$ at a finite value of $T$.
On the other hand, if $X=0$, which represents a model with no late time
inflation,  the limiting value of $Y$ is model dependent.

\section{Summary and Conclusions}\la{secC}

We have analyzed the tachyon field motivated by brane-antibrane interaction
with an effective BI type lagrangian with potential $V(T)$.
We have deduced the mass of the tachyon field $T$ as the pole of the propagator which
is not given   by the second derivative of $V$ nor by the second
derivative of the effective potential   $F \propto Log[V]$. The mass of
$T$ is given by eq.(\ref{mT}).

Since $T$ does not have canonical kinetic terms we have generalized the
commonly used transformation $\varphi=\int \sqrt{V} dT$, which is only
valid for small $\dot T$, to the transformation $\phi=\int (\rho^{3/2}/\sqrt{V}) dT$
valid for all $\dot T$ and which reduced to the transformation of $\varphi$
in the limit of small $\dot T$. We have derived  the equation of motion
for $\phi$ and determined the effective potential $U(\phi)$ given by
eq.(\ref{U}) and  its  mass eq.(\ref{mp}). We have shown that
the normalized tachyon field $\phi$, in a Minkowski metric, is defined only in the region between the
zeros of the original $V(T)$ potential
and the field $\phi$ takes the whole range between $-\infty$ and
$\infty$.

We have used the simple but powerful tool of L'H\^{o}pital rule to derive exact
slow roll conditions for late time inflation, i.e. dark energy.
As a side result we derive the potential which leads to a scaling behavior
for a standard scalar field and for a tachyon field.

\begin{acknowledgments}

A.M. would like to  thank the kind hospitality of
the DAMTP of the University of Cambridge and specially
F. Quevedo during his short
visit and the Royal-Society and the Academia Mexicana
de Ciencias for financial support. A.M. would
like to thank
F. Quevedo   for many
useful discussions and early collaboration of this manuscript.
A.M. would also like to thank the  This work was also supported in
part by CONACYT project 45178-F and DGAPA, UNAM project
IN114903-3.

\end{acknowledgments}

\end{document}